\documentclass[aps,showpacs,amsmath,amssymb,nofootinbib,preprintnumbers]{revtex4}
\usepackage{graphicx}
\usepackage{color}

\def \be  {\begin{equation}}
\def \ee  {\end{equation}}
\def \bea {\begin{eqnarray}}
\def \eea {\end{eqnarray}}

\begin{document}

\preprint{ECTP-2010-09}

\title{Acceleration and Particle Field Interactions of Cosmic Rays II: Calculations}
\author{A.~Tawfik}
\email{drtawfik@mti.edu.eg}
\affiliation{Egyptian Center for Theoretical Physics (ECTP), MTI University,
Cairo-Egypt}
\author{A.~Saleh}
\affiliation{Egyptian Center for Theoretical Physics (ECTP), MTI University,
Cairo-Egypt}
 \author{M.~T. Ghoneim}
\affiliation{Physics Department, Faculty of Science, Cairo University, Cairo-Egypt}
\author{A.~A. Hady}
\affiliation{Astrophysics Department, Faculty of Science, Cairo University, Cairo-Egypt}

\date{\today}

\begin{abstract}

Based on the generic acceleration model, which suggests different types of electromagnetic interactions between the cosmic charged particles and the different configurations  of the electromagnetic (plasma) fields, the ultra high energy cosmic rays are studied. The plasma fields are assumed to vary, spatially and temporally. The well-known Fermi accelerations are excluded. Seeking for simplicity, it is assumed that the energy loss due to different physical processes is negligibly small. The energy available to the plasma sector is calculated in four types of electromagnetic fields. It has been found that the drift in a time--varying magnetic field is extremely energetic. The energy scale widely exceeds the Greisen-Zatsepin-Kuzmin (GZK) cutoff. The polarization drift in a time--varying electric field is also able to raise the energy of cosmic rays to an extreme value. It can be compared with the Hillas mechanism. The drift in a spatially--varying magnetic field is almost as strong as the polarization drift. The curvature drift in a non--uniform  magnetic field and a vanishing electric field is very weak.  

\end{abstract}


\maketitle


\section{Introduction}

The Fermi acceleration mechanisms \cite{fermi49,shockB} have been extended to include different types of electromagnetic (plasma) interactions \cite{Tawfik:2010jh}. In present work, the total energy gained by cosmic charged particles is calculated according the generic acceleration model \cite{Tawfik:2010jh}. The arguments given in Ref. \cite{Tawfik:2010jh} on introducing non--relativistic formalism, are respected in the present work. It is aimed to show that the plasma field interactions, which originally are not implemented in the Fermi mechanisms (first-- and second--order), can contribute with a considerable amount to the ultra high energy of cosmic rays (UHECR). In the generic acceleration model \cite{Tawfik:2010jh}, the astrophysical objects, in which one or more types of plasma fields have been observed and/or predicted, are taken into consideration. On the other hand, the energy loss due to the different physical processes such as synchrotron, curvature and Bremsstrahlung radiation, etc. is entirely disregarded. The electric potential difference and $\mathbf{E}\times\mathbf{B}$, polarization, curvature and gravitational drifts are conjectured as sources for accelerating the cosmic charged particles by plasma field interactions. The plasma fields themselves can be homogeneous, non--homogeneous or time--spatially varying. It is obvious that the numerical calculations offer a tool to judge about the most effective mechanism, which likely derives the cosmic charged articles to an extreme energy value. 

The origin of UHECR is one of the greatest mysteries in modern physics. In order to compare their energy scale, let us recall that the current acceleration technologies are merely able to come up with ${\cal O}(10^{12})\,$eV. On the other hand, several observatories around the world have registered cosmic rays with primary energies $>10^{20}~$eV~\cite{pap0,pap0b,pap1,pap2,pap3,pap4}. From the theoretical point of view, the protons, which are the dominant components of UHECR, would posses energy values much higher than the so-called Greisen-Zatsepiti-Kuzmin (GZK) cutoff $10^{19.7}~$eV. They are conjectured to suddenly lose their energies in photopion production processes with the cosmic microwave background radiation \cite{pap5,pap6,pap7}. 

When a charged particle moves with reference to an equipotential surface where the electric potential $\mathbf{\phi}$ is constant and the existing electric field is stationary, then the equipotential surface is considered not to be coincide with the magnetic surface. In other words, the components of electric $\mathbf{E}$ and magnetic field strength $\mathbf{B}$ are perpendicular to each other. This leads to an $\mathbf{E}\times\mathbf{B}$ drift with velocity $\mathbf{V}_\perp$ and an energy ${\cal E}=q\,\mathbf{E}\cdot\mathbf{r}$, where $q$ is the static electric charge and $\mathbf{r}$ is the distance covered by the accelerated particle. This idea has been proposed by Hillas \cite{hillas1},  long time ago. In the generic acceleration model, this basic idea has been rather generalized. The configurations of the electromagnetic fields are designed to cover all possibilities, including the one in which the equipotential surface coincides with the magnetic surface.  

According to the generic model, the energy gained by cosmic charged particles along their path through a decreasing electric potential in a totally conserved field with varying spatial intensities is given as $\mathbf{{\cal E}}=2q\mathbf{E}\cdot\mathbf{r}$.
That this model predicts two times the Hillas energy can be understood according to the configuration of the fields and their variation in spatial and/or temporal dimensions.  

It is conjectured that the magnetic fields of the cosmic plasmas likely verify the so-called frozen-in condition which means that the magnetic flux $\Psi=B\,R^2$ is globally conserved \cite{pap0a}. Both electric and magnetic fields are correlated. They move coincidentally. On the other hand, the ideal magneto-hydro-dynamic (MHD) plasmas is conjectured to satisfy the condition $\mathbf{E}\cdot\mathbf{B}=0$ everywhere. They ensure that the generalized Ohm's law $\mathbf{E}=-\mathbf{V}\times\mathbf{B}$. The well--known Hillas criterion essentially relies on both conditions and leads to the non--relativistic relation ${\cal E}=q\,V\,B\,R$, where $R$ is the characteristic size of linear accelerator and $V$ is the velocity of the background plasmas. 

Hillas graph orders the magnetic field strength $\mathbf{B}$ versus the linear size $R$ for various astrophysical objects. It is found that the total energy, which would be available to the cosmic charged particles from their interactions with the magnetic fields described in Hillas model, can be as large as $10^{20}\,$eV \cite{pap1a}. In the generic model \cite{Tawfik:2010jh}, the configurations of $\mathbf{E}$ and $\mathbf{B}$ are generalized. This results in two components. First one is suppose to be responsible for the gyration perpendicular to both electric and magnetic fields. The other one is the accelerating component. Directions of both components, acceleration and gyration are coincident \cite{Tawfik:2010jh}. In present work, we introduce numerical calculations for the energy gained in these fields. 

Alternative hypotheses about the acceleration through electrostatic potential difference have been modelled for various astrophysical objects \cite{esp1}. Neutron stars (like pulsars, magnetars, etc.) are well known examples of astrophysical objects, in which electrostatic potential might be available. From theoretical point of view, standard pulsar magnetospheric, polar cap and slot gap models have been utilized \cite{esp1} in order to explain the ultra high energy based on electrostatic acceleration. Accordingly, the electric potential difference between center and edge of the polar cap is given as $\Delta\mathbf{\phi}\approx \Omega B_s r^2/2c$, where $r$ is the radius, $\Omega$ is the angular velocity and $B_s$ stands for magnetic field strength at the surface. Such a result can also be implemented in the generic model. The correlations between coordinates of UHECR and stellar objects can be taken into consideration as well. This would offer an alternative method to find out additional sources for UHECR and would be a next extension to the generic acceleration model.

The present paper is organized as follows. In the section that follows, \ref{sec:em1}, the energy gained by cosmic charged particles in various electromagnetic fields is calculated. Examples for the drift energies arising in four types of  electromagnetic fields are elaborated as follows. The drift of spatially varying magnetic field is given in section \ref{sec:dBdx}. Section \ref{sec:curv} is devoted to the curvature drift arising in a non--uniform magnetic field and a vanishing electric field. The polarization drift of the time--varying electric field is calculated in section \ref{pol-sec}. Last but not least, the time--varying magnetic field has been utilized to calculate the difference of drift energy in section \ref{sec:dBdt}. The discussion and final conclusions are given in section \ref{sec:dc}. This might explain the extreme high energy that the cosmic charged particles would gains in these types of electromagnetic fields.

\section{Drift energies in various electromagnetic fields}
\label{sec:em1}

In this section, we calculate the energy that the cosmic charged particles would gain through their interactions with the different types of the plasma fields \cite{Tawfik:2010jh}. As discussed in the introduction, the generic acceleration model offers a complete set of extensions to Hillas idea, where $\mathbf{E}\times\mathbf{B}$ drifts have been calculated, exclusively. It is worthwhile to remember again, that these field interactions would come up with a considerable part, may be a very large portion, of the total energy. Fermi acceleration mechanisms, which have been intensively studied over the last few decades, contribute with another considerable part. The energy loss due to various physical processes would reduce the total energy with a non--negligible amount. Since these processes would be very dominant under very special conditions and certainly rely on a very rich physics, we therefore like to study them quantitatively in a future work. Seeking for simplicity, we assume that such physical processes are very weak to be neglected.

\subsection{Spatially varying magnetic field $B$}
\label{sec:dBdx}

\begin{table}[htp]
\begin{center}
 \begin{tabular}{|c | c | c | c |}\hline\hline
  Object   &  R [m]     &  B [Tesla]  &  {$\cal E$} [eV] \\\hline\hline
Neutron Stars & $10^{4}$   & $10^{5}-10^{8}$  & $1.6 \cdot 10^{16}-1.6\cdot 10^{19}$ \\ \hline
Magnetars & $10^{4}$   & $10^{8}-10^{11}$  & $1.6 \cdot 10^{19}-1.6\cdot 10^{22}$ \\
\hline
Pulsars & $10^{4}$   & $10^{9}$  & $1.6 \cdot 10^{20}$ \\ \hline
Pulsar wind & $3\cdot10^{13}$  & $10^{-9}$  & $0.5 \cdot 10^{12}$ \\ \hline
SNR & $3\cdot10^{17}$ & $10^{-10}$  & $0.5 \cdot 10^{15}$ \\
\hline
Normal Stars & $10^{8}$ & $10^{-7}-10^{-1}$ & $1.6 \cdot 10^{8}-1.6\cdot 10^{14}$ \\
\hline
AGN & $10^{13}$ & $10^{-4}$  & $1.6 \cdot 10^{16}$ \\ \hline
AGN (Nucleus)& $3\cdot 10^{12}$ & $10^{-1}$  & $0.5 \cdot 10^{19}$ \\ \hline
AGN (Related Jets)& $3\cdot 10^{13}$ & $10^{-4}$  & $0.5 \cdot 10^{17}$ \\ \hline
AGN (Hot Spots)& $3\cdot 10^{19}$ & $10^{-8}$  & $0.5 \cdot 10^{19}$ \\ \hline
AGN (Radio Lobes)& $3\cdot 10^{22}$ & $10^{-10}$  & $0.5 \cdot 10^{20}$ \\ \hline
Galactic Disk & $10^{18}$ & $10^{-10}-10^{-9}$  & $1.6 \cdot 10^{15}-1.6\cdot 10^{16}$ \\
\hline
Galactic Cluster & $3\cdot10^{23}$ & $10^{-11}$  & $0.5 \cdot 10^{20}$ \\
\hline
GRB & $3\cdot10^{11}$ & $10^{-1}$  & $0.5 \cdot 10^{18}$ \\
\hline
Cen--A & $2.5\cdot10^{20}$ & $10^{-9}$  & $0.4 \cdot 10^{19}$ \\ \hline
 \end{tabular}
\caption{Expansions and magnetic field strengths of different astrophysical objects are used to calculate the energy the the cosmic charged particles would gain from the plasma field interactions. In these calculations, the velocity $v$ is assumed to be non--relativistic; $v\sim 0.1 c$ and the cosmic charged particles  are assumed to curry a unit charge, $q=1.6\cdot 10^{-19}\,$C. The last line is added to compare with ordinary astrophysical objects with the Cen-A. }
\label{tab-1}
\end{center}
\end{table}

The gradient drift is associated with a non--homogeneous magnetic field $\mathbf{B}$ and a vanishing electric field, $\mathbf{E}=0$. We assume a very simple case of a spatially-varying magnetic field. A particle of mass $m$ and static electric charge $q$ moving with velocity $v$ in such magnetic field would be accelerated to an average energy 
\be\label{E1}
\left\langle {\cal E} \right\rangle = q\, \frac{v\, r_l}{2}\cdot{r}\frac{d\mathbf{B_z}}{d\mathit{y}},
\ee
where the variation of this magnetic field is assumed to be in $z-$direction and 
\bea\label{eq:dbz1}
\frac{\mathbf{dB_z}}{d\mathbf{y}}&\approx& \frac{B_z}{L} \ll \frac{B_z}{r_l}.
\eea
$L$ is the characteristic size of the acceleration area. $r_l$ is defined as the Larmor radius. In order to assure that the particle does not exceed the horizon of the acceleration area, the inequality given in Eq. (\ref{eq:dbz1}) has to be valid, i.e. $L$ has to be much greater than $r_l$. Let us assume that the cosmic charged particle gyrates along $z$ and $y$ directions and covers a distance $r_l \approx L$. The last assumption is essential to assure that particle is about to leave the acceleration area. Thus
\be\label{E3}
  \left\langle{\cal E}\right\rangle\approx q\,\frac{v\,L}{2}\,B_0,
\ee
where $B_0$ is the surface magnetic field strength of the stellar object.

To answer the question about the nature of stellar objects, in which such a non-uniform magnetic field would be possible, we suggest the neutron stars which inhibit a very strong magnetic field strength spatially ranging from $10^9$ to $10^{12}\,$G depending on the radius or distance covered by the cosmic particle inside the magnetic field. Crab-pulsar and magnetars are well-known examples of stellar objects with extreme magnetic fields. The time period of their spinning rotation apparently affect the strength of the magnetic fields. Younger neutron stars would have a very much strong magnetic field and even a shorter time period of spinning $(33\,$ms) than the elder ones \cite{nstars}. The structure formation shocks \cite{shocks} are conjectured as possible sources. Also, spinning black holes have been suggested \cite{BH-accel} to accelerate the cosmic charged particle to ultra high energies, Tab. \ref{tab-1}. 

Another method to estimate an arbitrary Larmor radius $r_l$ would be its relation with the magnetic field strength $\mathbf{B}$ \cite{Tawfik:2010jh}. Then, the energy drift in non-uniform magnetic fields of neutron stars would be as much as $\sim4.9\times10^{16}\,$eV, which is apparently too small to assure ultra high energy to the cosmic charged particles. In Tab. \ref{tab-1}, we list the energies gained by cosmic charged particle in different astrophysical objects.

Such a result seems to verify Hillas model, which assumes that the energy gained by cosmic charged particles stems from a special type of drifts, explicitly. This is the polarization drift as will be given in section \ref{pol-sec}. The generic acceleration model \cite{Tawfik:2010jh} assumes that this energy is to be considered as one source. The Fermi accelerations and other types of field interactions are other sources.


\subsection{Non--uniform  magnetic field $B$ and vanishing electric field $E$ (curvature drift)}
\label{sec:curv}

We assume that the charged particle moves parallel to the field lines. The latter are in turn assumed to be curved. Therefore, the particle experiences a centrifugal force. Under the assumption that the electric field $\mathbf{E}$ vanishes and the magnetic field $\mathbf{B}$ being {\it time-independent} and {\it weakly inhomogeneous}, then the charged particle would gyrate around the magnetic field lines describing a helical path with Larmor radius $\mathbf{r}_l$ \cite{Tawfik:2010jh}. Under these conditions, we get
\be\label{cen1}
\mathbf{F}_R = m\,v^2_\parallel\,\frac{\mathbf{R}_c}{R^2_c},
\ee 
where $\mathbf{R}_c$ is the local radius of curvature. Inserting the equation of general force drift
\bea
\mathbf{v}_d &=& \frac{1}{q}\frac{\mathbf{F}\times \mathbf{B}}{B^2}
\eea
into Eq.\,(\ref{cen1}) leads to an expression for the {\it curvature} drift
\be\label{cen2}
\mathbf{v}_R = \frac{m\,v^2_\parallel}{q}\,\frac{\mathbf{R}_c\times\mathbf{B}}{R^2_c\,B^2}.
\ee 
It is obvious that $\mathbf{v}_R$ is a constant velocity. According to Eq. (\ref{cen2}), the acceleration vanishes. Therefore,
the curvature drift - in non--relativistic limit - seems to be proportional to the parallel component of the particle energy; ${\cal E}_\parallel=\frac{1}{2} m  v^2_\parallel$. In relativistic limit, this component can be written as 
\be\label{cen3}
{\cal E_\parallel} = \gamma m_o\,c^2.
\ee
It is apparent, that the energy of Eq. (\ref{cen3}) does not depend on the astrophysical parameters such as electric field $\mathbf{E}$, magnetic field $\mathbf{B}$, expansion $\mathbf{r}$, etc. The maximum energy value would be of the order of ${\cal O}(10^6)\,$eV.


\subsection{Time--varying electric field $E$ (polarization drift)}
\label{pol-sec}

Assuming that the magnetic field strength $B$ is homogeneous and non--varying while the electric field strength $\mathbf{E}$ is temporarily varying within the acceleration area. At a particular distance $\mathbf{r}$, the acceleration and energy are given as follows.
\bea
\mathbf{a} &=& \frac{m}{q B^2} \left(-\omega^2 \mathbf{E}(\mathbf{r},t)\right), \label{pol-a} \\
\mathbf{E}(\mathbf{r},t) &=& E_0 \exp(i\omega t) \label{pol-E},
\eea 
where angular velocity $\omega$ is related to the radius $\mathbf{r}_l$. Utilizing Ohm's law in MHD--plasma and $\mathbf{E}=\mathbf{V}\times \mathbf{B}$, then the acceleration in Eq. (\ref{pol-a}) can be rewritten as, 
\bea
\mathbf{a} &=& \frac{q}{m} \left(-\mathbf{V}\times \mathbf{B}\right) \label{pol-a2}.
\eea
Assuming a special case, in which both $\mathbf{V}$ and $\mathbf{B}$ are perpendicular to each other, then the polarization energy drift simply reads
\bea \label{pol-Espl}
{\cal E} &=& q\,\mathbf{V}\,\mathbf{B}\,\mathbf{R},
\eea 
where $\mathbf{R}$ and $\mathbf{V}$ being the linear size of the astrophysical object and velocity of the background plasma, respectively. Equation (\ref{pol-Espl}) seems to remind with the Hillas energy. It is notable that the polarization drift of a charged particle looks similar to Hillas model. It is obvious that the latter is nothing but twice the energy gained by the {\it same} cosmic charged particle in a spatially varying $\mathbf{B}$ (section \ref{sec:dBdx}).


\subsection{Time--varying magnetic field $\mathbf{B}$ (grad $\mathbf{E}$ drift)}
\label{sec:dBdt}

Assuming that the magnetic field $\mathbf{B}$ is not constant over the entire time interval $t$, then we can apply Faraday's law so that the time--varying magnetic field strength, $d\mathbf{B}/dt$, turns to be given by $-\nabla\times\mathbf{E}$. Such a relation apparently implies that the time--varying $\mathbf{B}$ generates a ''grad $\mathbf{E}$'' drift. The latter raises the acceleration so that the change in the energy reads  
\bea
\delta \mathbf{E}_{\nabla E} &=& 2\pi \frac{\mu}{\Omega_c} \frac{d \mathbf{B}(t)}{d t},
\eea
where $\mu$ is the magnetic moment. It is given by $m v_{\perp}^2/2\mathbf{B}$. $\Omega_c$ is the cyclotron frequency, which is equivalent to $q\mathbf{B}/m$. It is obvious that this drift strongly depends on the particle's mass, $m$, and magnetic field strength $\mathbf{B}_0$. 
\bea
\delta \mathbf{E}_{\nabla E} &=& \frac{\pi}{q}\,\left(\frac{m v_{\perp}}{B_0}\right)^2 \frac{d \mathbf{B}(t)}{d t},
\eea

Again, the crab pulsars and/or neutron stars would be subject to this drift. In these astrophysical objects, phases of magnetic growth have been conjectured. Many authors \cite{ns1,ns2} have suggested that the magnetic fields in neutron stars, for instance, can be explained by the so--called thermally driven battery mechanism. The latter means that the neutron starts would posses a magnetic field of $\leq 10^{8}\,$G at their birth and over a period of $\sim 10^{5}\,$y this magnetic field raises to $\sim 10^{12}\,$G. Other mechanisms which temporarily affect the magnetic field include the rotation period, matter transfer phase and periods of spin--up and -down, especially, in the dipole fields. In the growth phase, the temporarily growing magnetic field strength $\mathbf{B}(t)$ and spindown age $t_s(t)$, respectively, would read \cite{crab}
\bea
\mathbf{B}(t) &=& \mathbf{B}_0\, \exp(t/\tau), \label{bt_crabP}\\
t_s(t) &=& t_{s0}\exp(-2t/\tau) + \frac{\tau}{2}[1-\exp(-2t/\tau)], \label{ts_crabP}
\eea
where $\tau$ stands for the characteristic timescale for the growth of the magnetic field. Equations (\ref{bt_crabP}) and (\ref{ts_crabP}) describe an exponential growth. Regardless the GZK-cutoff \cite{pap5,pap6}, a simple calculation shows that this magnetic field would be able to accelerate the proton up to $\sim 10^{22}\,$eV. This huge energy value would be available to protons which are positioned in this rapidly changing magnetic field and after the acceleration they will able to fly away from the accelerating stars. The well--known Fermi mechanisms are not taken into account. Also, the energy loss is not calculated. 

Few remarks on the origin of magnetic fields in pulsars and magnetars are now in order. there are three scenarios, 1) generation from fossil fields, 2) internal generation in progenitor, and 3) generation during core collapse \cite{sourc}. The third scenario is very much interesting. Tidal flows and deformations have been suggested to explain the generation of magnetic fields in binary celestial systems. These processes are reversible in a short period \cite{dolg}. Also, merging of a binary neutron star system has been suggested to produce a extremely rapid mechanism for ultra-strong magnetic field \cite{pricce}. This has been suggested to be responsible for the short class of gamma ray bursts (GRB). MHD fields are amplified via Kelvin--
Helmholtz instabilities beyond magnetar field strength. They might represent
the strongest magnetic fields in the universe. The amplification seem to occur
in the shear layer positioned between the binary stars and at a time scale
of $\sim1\,\mu$s. Therefore, the energy that a singly charged particle, like proton, would gain seems to be incredibly high.


\section{Discussion and Conclusions}
\label{sec:dc}

The energy that cosmic charged particles would gain from their interactions with different types of electromagnetic fields has been calculated. It has been found that the energy drifts can be as large as the GZK-cutoff \cite{pap5,pap6}, although Fermi mechanisms are excluded. It is essential to mention here that the energy loss due to various physical processes has been disregarded, as well. They would be very dominant under very special conditions and certainly rely on a very rich physics. Seeking for simplicity, we assume that such the energy--reducing processes are very weak to be neglected. Nevertheless, a quantitative estimation will be elaborated in a future work.

In present work, four types of plasma drifts have been calculated for a particle of mass $m$ and static electric charge $q$. The particle initial velocity, either finite or vanishing $v$, shall be increased over its path through the plasma field. The utilized model \cite{Tawfik:2010jh} assumes general configurations for electric $\mathbf{E}$ and magnetic field strength $\mathbf{B}$. Such an arrangement results in two components. The first one is responsible for the gyration perpendicular to both electric and magnetic fields. The other one is responsible for the acceleration. It has been found that the directions of both components, acceleration and gyration are coincident.  

The spatially varying magnetic field $B$ seems to verify Hillas model. The latter assumes that the energy gained by cosmic charged particles stems from a special type of drifts, explicitly. This is the polarization drift. The generic acceleration model \cite{Tawfik:2010jh} assumes that this energy is to be considered as one source. The Fermi accelerations and other types of field interactions are other sources.  

The neutron stars, especially the younger ones, inhibit very strong magnetic fields, which seems to vary from $10^9$ to $10^{12}\,$G depending on the radius or distance covered by the accelerated particle. The crab-pulsar and magnetars are stellar objects with extreme magnetic fields. The time period of their spinning rotation apparently affect the strength of magnetic fields. The spinning periods of younger neutron stars secure extreme magnetic fields. The structure formation shocks and spinning black holes can accelerate cosmic charged particle to ultra high energies. 

The drift of a non--uniform magnetic field $\mathbf{B}$ and a vanishing electric field $\mathbf{E}$ seems to be very weak, so that the maximum energy is just couple MeV. Apparently, this kind of drifts is responsible for curving the accelerated particle around the non--uniform magnetic field $\mathbf{B}$. Therefore, it would feed the final energy with a small amount. Even in the relativistic limit, the contribution of the curvature drift is minimum.

The drift in a time--varying electric field $E$ gives an energy value, which is related to magnetic field strength. It reflecting the Ohm's law and the configurations of the two fields ($\mathbf{E}$ and $\mathbf{B}$). In MHD plasma, $\mathbf{E}$ and $\mathbf{B}$ are perpendicular to each other. Therefore, the relation between them is governed by the Ohm's law. As discussed above, this {\it polarization} drift results in the well--known Hillas energy scale.  

Last but not least, the drift in a time--varying magnetic field $\mathbf{B}$, so--called ''grad $\mathbf{E}$'' drift, seems to be very much effective. It relies on Faraday's law. On the other hand, the final energy that the cosmic charged particle would gain in this field is directly related to the time--varying magnetic field strength. Crab pulsars and neutron stars are conjectured to have phases, in which the magnetic field rapidly change with the time. Accordingly, their fields have been explained by thermally driven battery mechanism. Rotation, matter transfer, spin--up and -down periods, especially, in the dipole fields, are examples of additional mechanisms which temporarily affect the magnetic field. 
Regardless the conservation of GZK-cutoff \cite{pap5,pap6}, such a magnetic field would be able to accelerate the proton, for instance, up to $\sim 10^{22}\,$eV. In this huge energy value, the well--known Fermi mechanisms are not taken into account.   

So far, we conclude that the ''grad $\mathbf{E}$'' drift (in time--varying magnetic field $\mathbf{B}$) is very much powerful. It gives the largest energy value. Also, the polarization drift (in a time--varying electric field $\mathbf{E}$) is able to raise the cosmic ray energy to an extreme value. It can be compared with the Hillas mechanism. The drift in a spatially--varying magnetic field $\mathbf{B}$ is almost as strong as the polarization drift. The fourth drift, the curvature a non--uniform  magnetic field and a vanishing electric field, is very weak.  


\end{document}